\documentclass[12pt]{iopart}
\usepackage{iopams}
\usepackage{graphicx, epsfig}

\newcommand{\apjl}{{Astrophys.~J.~Lett.}}

\def\fun#1#2{\lower3.6pt\vbox{\baselineskip0pt\lineskip.9pt
        \ialign{$\mathsurround=0pt#1\hfill##\hfil$\crcr#2\crcr\sim\crcr}}}

\def\dslash{\not{\hbox{\kern-2pt $\partial$}}}
\def\Dslash{\not{\hbox{\kern-4pt $D$}}}
\def\Oslash{\not{\hbox{\kern-4pt $O$}}}
\def\Qslash{\not{\hbox{\kern-4pt $Q$}}}
\def\pslash{\not{\hbox{\kern-2.3pt $p$}}}
\def\kslash{\not{\hbox{\kern-2.3pt $k$}}}
\def\qslash{\not{\hbox{\kern-2.3pt $q$}}}

 \newtoks\slashfraction
 \slashfraction={.13}
 \def\slash#1{\setbox0\hbox{$ #1 $}
 \setbox0\hbox to \the\slashfraction\wd0{\hss \box0}/\box0 }

\def\ee{\end{equation}}
\def\be{\begin{equation}}

\begin{document}
\title{Improved Cosmological Parameters 
Constraints 
from CMB and H(z) Data} 
\author{Daniel G. Figueroa$^{1}$, Licia Verde$^{2,3}$ and  Raul Jimenez$^{2,3}$\\
{\it $^1$Instituto de F\'isica Te\'orica UAM/CSIC, Facultad de Ciencias,
Universidad Aut\'onoma de Madrid, Cantoblanco, Madrid 28049, Spain.\\
$^2$ ICREA \& Institute of Space Sciences (CSIC-IEEC), Campus UAB, Bellaterra, Spain.\\
$^3$ Dept. of Astrophysical Sciences, Princeton University, Princeton,  
NJ08544, USA\\
}}

\begin{abstract}

We discuss the cosmological degeneracy between the Hubble parameter $H(z)$, the age of the universe and cosmological parameters describing simple variations from the minimal $\Lambda$CDM model. We show that independent determinations of the Hubble parameter $H(z)$ such 
as those recently obtained from ages of passively evolving galaxies
, combined with Cosmic Microwave Background data (WMAP 5-years), provides stringent constraints on possible deviations from the $\Lambda$CDM model. In particular we find that this data combination constrains at the 68\% (95\%) \textit{c.\,l.}\,\,the following parameters: sum of the neutrino masses $\sum m_\nu < 0.5\,(1.0)$ eV, number of relativistic neutrino 
species $N_{\rm rel} = 4.1^{+0.4}_{-0.9}\,(^{+1.1}_{-1.5})$, dark energy equation of state parameter $w = -0.95 \pm 0.17\,(\pm\,0.32)$, and curvature $\Omega_k = 0.002 \pm 0.006\,(\pm\,0.014)$\,, in excellent agreement with dataset combinations involving Cosmic Microwave Background, Supernovae and Baryon Acoustic Oscillations. This offers a valuable consistency check for systematic errors.
\end{abstract}



\maketitle

\section{Introduction}
The recent measurements of cosmic microwave background (CMB) anisotropies and 
polarization \cite{spergelwmap2,dunkleywmap08}, alone or in combination with other cosmological data sets,  have provided  confirmation of the standard
cosmological model and an accurate determination 
of some of its key parameters. 

In particular, the new determination
of the age of the Universe $13.68 \pm 0.13$ Gyrs improves by an
order of magnitude previous determinations from, e.g.,
cosmochronology  of long-lived radioactive nuclei \cite{thorium} and 
population synthesis of the oldest stellar populations \cite{Jimenez96,dunlop96,chaboyerkrauss03} and by a factor of 2 previous determinations from CMB data.

With cosmological parameters so tightly constrained within the framework of the standard flat-$\Lambda$CDM model, it is important however to constrain possible deviations from the standard cosmological model. Beyond the primordial parameters describing the shape of the primordial power spectrum and late-time parameters such as the optical depth to the last scattering surface,  
CMB observations so far constrain directly parameters such as \cite{kosowsky} the angular size distance to last scattering combined with the sound horizon at decoupling, the baryon-to-photon ratio and the redshift of matter radiation equality.
This implies that, for models beyond the standard flat-$\Lambda$CDM,  CMB data alone still show large degeneracies among ``derived"\footnote{Note that the name "derived parameters" has been sometimes used in the literature with a slightly different emphasis, denoting parameters such as the bias $b$, $\sigma_8$, etc..} cosmological parameters such as  the matter density parameter $\Omega_m$, the curvature $\Omega_k$,  the dark energy equation of state paramenter $w$,  the  effective number of relativistic neutrino species $N_{\rm eff}$,  the sum of neutrino masses $\sum m_{\nu}$ and  the Hubble parameter  $H_0$.
For example  (see e.g., \cite{Eisensteinwhite04, debernardisetal08, de Bernardis:2007bu}),  departures from the standard model described by a deviation from  $3$ neutrino species, can arise from  the decay of dark matter
particles \cite{bonometto98, lopez98,  hannestad98, kaplinghat01}, 
quintessence \cite{bean}, exotic models \cite{unparticles}, 
and additional hypothetical relativistic particles. This affects the matter-radiation equality yielding, even for a flat, cosmological constant-dominated  model,  a degeneracy between $N_{\rm eff}$, $H_0$ and  $\Omega_m$. 
A departure from dark energy  being described by a cosmological constant (i.e. a component with equation of state $w\neq-1$), yields a  different   angular size distance to last scattering, and thus degeneracy between $w$, $H_0$ and $\Omega_m$ even for a flat universe. Finally, relaxing the flatness assumption yields the so-called ``geometric degeneracy" (between age or $H_0$ and $\Omega_m$ and  $\Omega_{\Lambda}$).   

In order to go beyond the concordance $\Lambda$CDM model parameters determination, one needs extra data sets that probe different physics and are affected by different systematics. In this work we concentrate on the measurements of $H(z)$ using passively evolving red-envelope galaxies and how using them helps to constrain cosmological parameters dropping the assumption of the concordance model.  In particular we show that the recent determinations of $H_0$ from the
HST key project \cite{hstkey} and $H(z)$ 
provided by \cite{svj} (SVJ; based on  
\cite{data} and references therein) can provide, when combined 
with CMB and other cosmological data, new and tighter constraints on deviations from the standard $\Lambda$CDM model, as first shown in \cite{de Bernardis:2007bu, data}. This approach  of combining different data-sets to constrain parameters that are otherwise poorly constrained,  is called ``concordance approach".  While it is a very powerful approach, the same ``concordance" approach is used to test data sets for systematic errors.   It is therefore important to 
consider enough data sets to  have an  over-constrained  problem and as diverse data sets as possible, relying on different physics and affected by different systematics. Only in this case, if all data sets agree, one can be confident that the systematic errors are safely below the statistical errors and that the cosmological constraints are robust. 

After obtaining constraints on deviations from the simple $\Lambda$CDM model obtained with WMAP 5-years data and $H(z)$ measurements, we compare them with those obtained from the combination of WMAP 5-years with Baryon Acoustic Oscillations and Supernovae data.
We find good agreement between the two approaches. 
We conclude that any possible systematic effect in the non-CMB data sets is below the statistical errors, and that there is no evidence for a deviation from the 
flat-$\Lambda$CDM model, thus offering  support to the standard cosmological model. 

\section{Data Analysis: Method}
The method to extract cosmological parameters from the different datasets that we adopt is based on the publicly available Markov Chain Monte Carlo
package \texttt{cosmomc} \cite{Lewis:2002ah} and the sampling of the posterior distribution given by  Monte Carlo Markov Chains released with the WMAP 5-years data \cite{dunkleywmap08}. The  standard $\Lambda$CDM model is described by the following  set of cosmological parameters: 
the physical baryon and CDM densities, $\omega_b=\Omega_bh^2$ and 
$\omega_c=\Omega_ch^2$, the density parameter of dark energy $\Omega_{\Lambda}$, the scalar spectral index, $n_{s}$ and amplitude $A_s$,
and the optical depth to reionization, $\tau$\footnote{We marginalize over the SZ amplitude parameter as done by the WMAP team \cite{dunkleywmap08}}.  For all these parameters the chosen boundaries of the priors do not affect the cosmological constraints. We  consider deviations from this model described by the addition of a single extra parameter. The models which show a significant degeneracy between $H(z)$ and the additional parameter are:  models where we add the possibility of having an extra-background of relativistic particles (parametrized by $N_{\rm eff} \neq 3.04$), or where we fix the effective number of neutrinos to $N_{\rm eff} = 3.04$ but allow them to have significant non zero mass $\sum m_{\nu}\ne 0$, models where we consider the possibility of a (constant) dark energy equation of state $w \neq  -1$, and finally models with non-flat geometry $\Omega_k\ne 0$. 

We then study how determinations of the rate of expansion $H(z)$ can constrain these deviations. We consider the Hubble key project determination of the Hubble constant \cite{hstkey} (HST) and the determination of the redshift dependence of the Hubble parameter $H(z)$ from observations of passively evolving galaxies \cite{svj} (SVJ). This combination (WMAP5+HST+ages) is referred to as ``WMAP5+H".

Finally, we also consider a model which deviates from the standard  $\Lambda$CDM by the addition of two parameters: $\Omega_k$ and $w$, and investigate how the $H(z)$ data helps break the  CMB-only degeneracy.

To conclude, we compare these constraints to those obtained with the combination WMAP 5-years with Supernovae and Baryon Acoustic Oscillations (\cite{komatsuwmap08,percivaletal07}). This combination  is referred to as ``WMAP5+SN+BAO".

\subsection{$H(z)$ determination}
An important observable to constrain cosmological parameters is a direct measurement of the Hubble parameter $H(z) = \frac{\dot a}{a}$, as this measures directly the expansion rate of the universe at a given redshift. For example, $H(z)$ is a more direct measurement of the equation of state of dark energy than the angular diameter distance $d_A(z)$ or the luminosity distance $d_L(z)$. This is easy to see by recalling that, adopting a FRW metric, using Einstein's equations and considering a flat universe composed of matter and dark energy with equation of state $p_Q = w_Q(z) \rho_Q$,  $H^2 = H_0^2 [\rho_T(z)/\rho_T(0)]$ and thus
\begin{equation}
\frac{H(z)}{H_0} = (1+z)^{\frac{3}{2}}  \left [ \Omega_M(0) + \Omega_Q(0) \exp \left [ 
3 \int_0^z \frac{dz'}{1+z'} w_Q \right ] \right ]^{1/2},
\end{equation}
where the subscripts $Q$, $M$ and $T$ refer respectively to the dark energy, the matter, and  the total contents. The quantities $d_A(z)$ and $d_L(z)$ are related to $H(z)$ via  $d_A(z)(1+z)=d_L(z)/(1+z) = \int_0^zdz'/H(z)$. 

While some of the current constraints on the dark energy equation of state parameter $w_Q(z)$, are based on integrated measurements of $H(z)$ (like the angular-diameter distance), other observables have already provided direct measurements of $H(z)$, like the determination of the star populations of luminous red galaxies \cite{svj}. Other techniques that can provide a direct measurement of $H(z)$ are the power spectrum of the peculiar velocities, as measured, for example, by the KSZ effect \cite{CHM07,kosowskyKSZ} or the Baryonic Acoustic Oscillation (BAO) scale in the radial direction \cite{seo03,seo07}. The BAO technique has recently received renewed attention because its potential to provide a standard ruler at different redshifts, and because of its robustness to systematic effects; it is thus being considered a powerful method to determine the nature of dark energy.

\begin{figure}
\begin{center}
\hspace{-0.1cm}\includegraphics[width=8cm, height= 12cm, angle=-90]{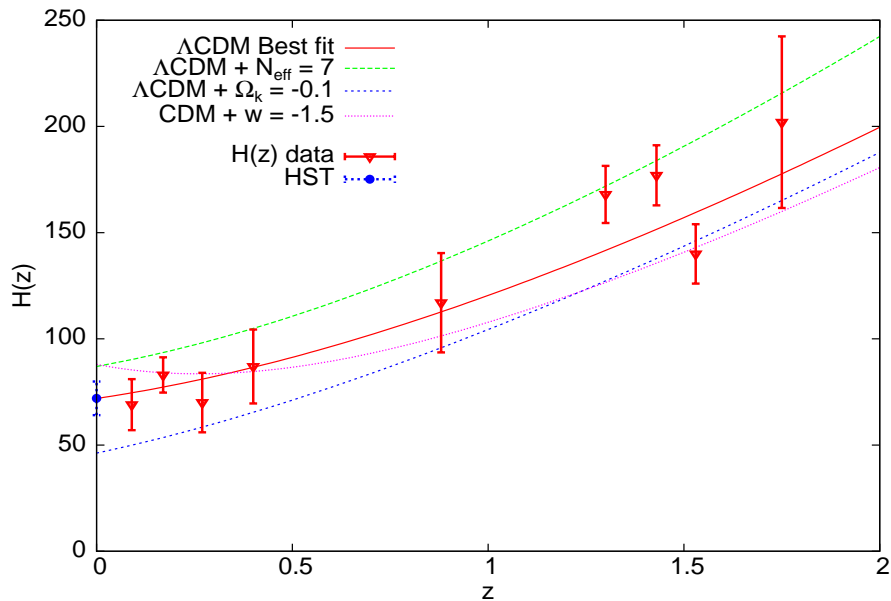}
\end{center}
\caption{The H(z) data and their error bars (from~\cite{svj}) used here.  
For comparison, we also plot $H(z)$ curves for several models compatible with CMB meauserements: the $\Lambda$CDM best WMAP fit (solid line) and the best WMAP fits to models which allow a number of relativistic species different than 3 (e.g. $N_{\rm eff}$ = 7, large dashed line), non-zero curvature (e.g. $\Omega_k = -0.1$, short dashed line) and finally Dark Energy with $w \neq -1.$ (e.g. $w = -1.5$, dotted line).}
\label{fig:H(z)}
\end{figure}

In this paper we constrain cosmological parameters for models beyond the standard $\Lambda$CDM, using the $H(z)$ determinations provided by \cite{svj}, obtained from the study of the evolution of the star populations in massive ($ > 2.2 L_*$) luminous red galaxies (LRG). Recent studies \cite{treu05,cool08} have clearly established that masive LRG have formed more than 95\% of their stars at redshifts higher than 4. These galaxies, therefore, form a very uniform population, whose stars are evolving passively after the very first short episode of active star formation \cite{heavens04,thomas05}. Because the stars evolve passively, these massive LRG are excellent cosmic clocks, i.e. they provide a direct measurement of $dt/dz$; the observational evidence discards further star formation activity in these galaxies. Dating of the stellar population can be achieved by modeling the integrated light of the stellar population using synthetic stellar population models, in a similar way to what is done for open and globular clusters in the Milky Way. The dating of the stellar population needs to be done on the integrated spectrum because individual stars are not resolved and therefore the requirements on the observed spectrum are stringent as one needs a very wide wavelength coverage, spectral resolution and very high signal-to-noise. In~\cite{cool08}, it has been shown that the spectra of these massive LRG at a redshift $z \sim 0.15$ are extremely similar, with differences of only 0.02 mag, which is another evidence of the uniformity of the stellar populations in these galaxies. There have already been examples of accurate dating of the stellar populations in LRGs \cite{svj,dunlop96,data,spinrad} where it has been shown that galaxy spectra with sufficient wavelength coverage (the UV region is crucial), wavelength resolution (about 3\AA) and enough $S/N$ (at least 10 per resolution element of 3\AA) can provide sensible constraints on cosmological parameters. The resulting $H(z)$ data obtained from these studies and used in this paper, are the points with error-bars shown in Figure~\ref{fig:H(z)}. The interested reader can find more details 
in \cite{svj,data}.

\begin{figure}[t]
\hspace{-0.55cm}\includegraphics[width=8cm, angle=0]{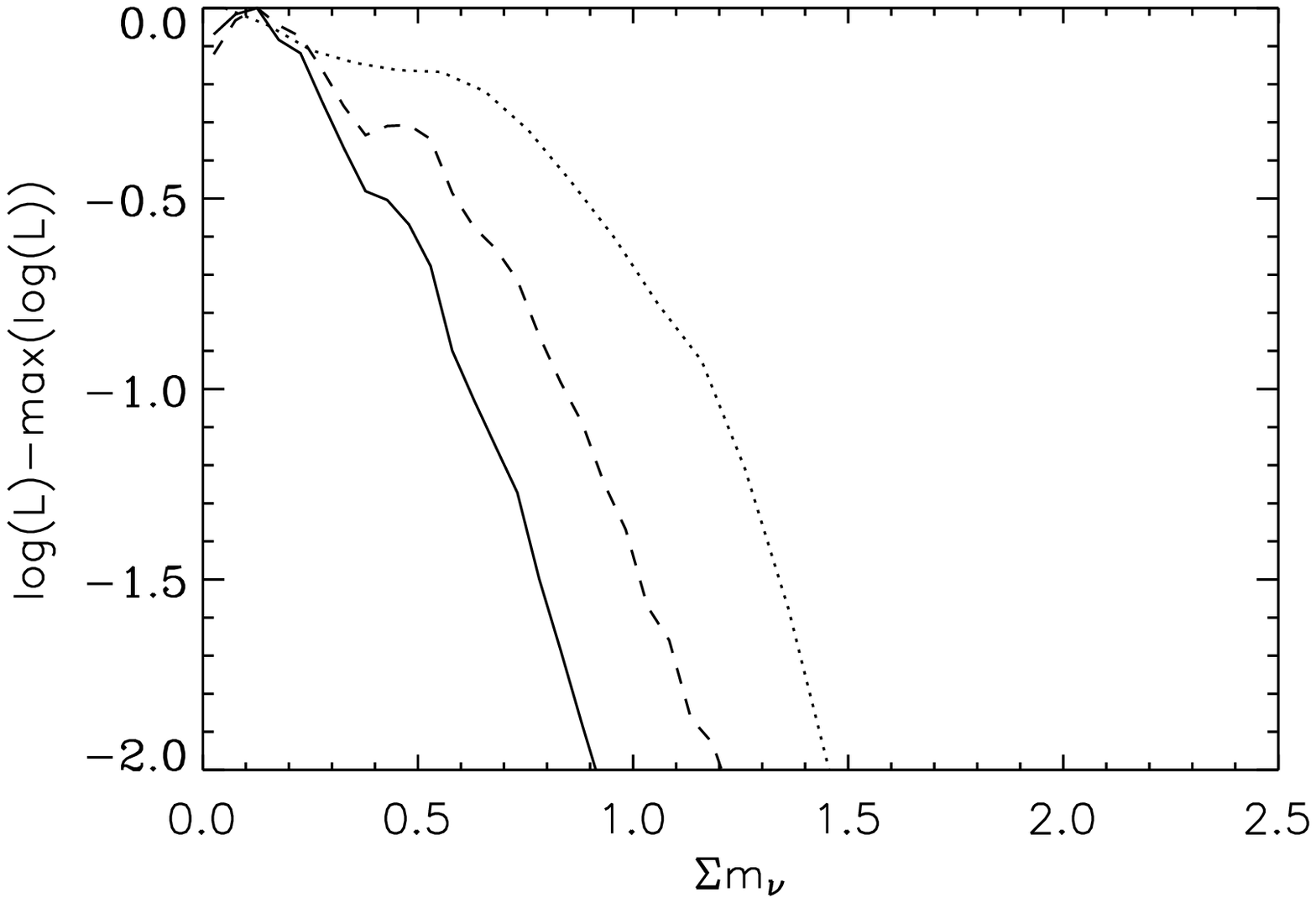}
\hspace{-0.1cm}\includegraphics[width=8cm, angle=0]{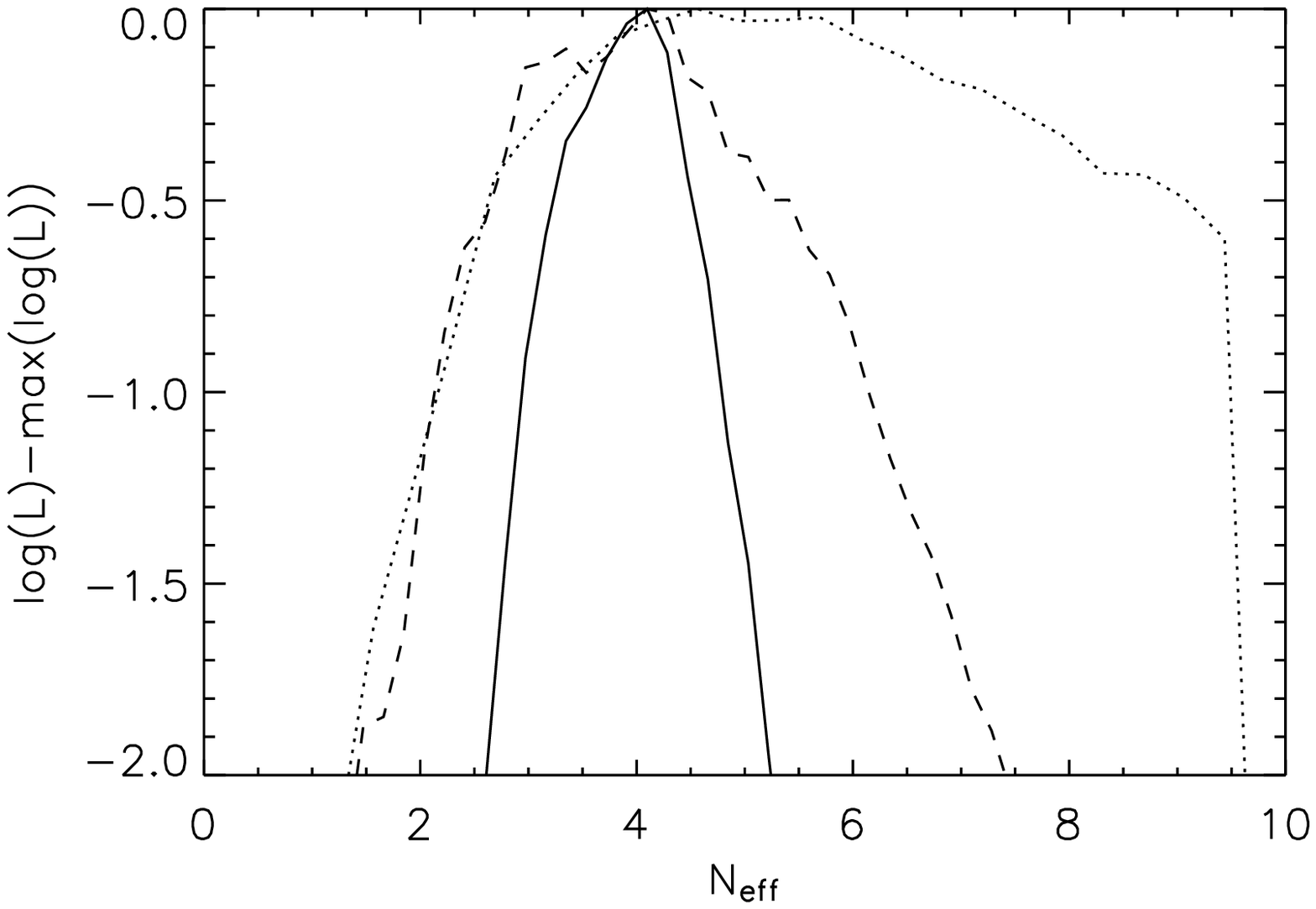}
\begin{center}

\caption{\textit{Left}: Constraints on the total mass of relativistic neutrinos from WMAP 5-years alone (dotted line), WMAP5+HST (dashed line) and WMAP5+H (solid line). The total sum of the neutrino masses, $\Sigma\, m_\nu$ is constrained to be below $0.48$ ($0.93$) eV at 68\% (95\%) confidence level, by the combination WMAP5+H. \textit{Right}: Constraints on the effective number of relativistic neutrinos species from WMAP 5-years alone (dotted line), WMAP5+HST (dashed line) and WMAP5+H (solid line). The effective number of neutrino species is constrained to be $N_{\rm rel} = 4.1^{+0.4}_{-0.9}\,(^{+1.1}_{-1.5})$ at the 68\% (95\%) confidence level. The WMAP 5-years only constraint has a hard prior $N_{\rm eff}<10$ imposed. Adding HST or H constraints make the determination insensitive to the prior. \label{lcdm_mnu}}
\end{center}
\end{figure}

\section{Results and conclusions}

Measurements of $H(z)$ constrain the age of the Universe at different redshifts and thus break the CMB-only degeneracy between the age of the universe today ($t_0$) and the parameters describing deviations from the $\Lambda$CDM model. As shown in table~\ref{tab:ages}, the age of the Universe as constrained by WMAP5-only data, is very sensitive to the presence of some of these parameters, especially to the possibility of having a background of $N_{\rm eff}$ relativistic particles (with $N_{\rm eff}$ not fixed to $3.04$) or allowing a non-zero curvature (see first row, 3rd and 5th columns of table~\ref{tab:ages}). 
This is because many models which deviates from the standard $\Lambda$CDM but are consistent with CMB data, are not a good fit to the $H(z)$ data. Some illustrative examples are shown in Figure 1.  
The combination WMAP5+H significatively reduces the degeneracy between $t_0$ and some of the ``extra" parameters, thus improving the constraints on the age of the Universe in those models, for example, by almost a factor of 3 for the case with $N_{\rm eff}$ not fixed to 3.04, and almost a factor of 5 for the case of non-zero curvature (see same columns but second row, in table~\ref{tab:ages}).

\begin{table}[htb]
\caption{\label{tab:ages} Determination of the age of the Universe ($68.3 \%$ c.l.) in several different cosmological models for WMAP 5-years data alone and WMAP5+H data. }
\begin{center}
\begin{tabular}{|c||c|c|c|c|c|}
\hline
AGE (Gyr)    & $\Lambda$CDM & $\Lambda$CDM+$N_{\rm eff}$ & $\Lambda$CDM +$\sum m_{\nu} $  & $\Lambda$CDM+$\Omega_k$ & wCDM \\
\hline
& & & & &\\
WMAP 5-years  & $13.69 \pm 0.13$ & $12.08 \pm 1.29$ & $14.06\pm 0.27$  &$16.32 \pm 1.76$   & $13.74 \pm 0.34$    \\
& & & & &\\
WMAP5+H  & $13.65^{+0.14}_{-0.10}$  &  $12.87^{+0.61}_{-0.31}$  &  $13.81^{+0.24}_{-0.14}$  &  $13.61^{+0.29}_{-0.44}$ &  $13.67^{+0.24}_{-0.08}$  \\
& & & & &\\
\hline
\end{tabular}
\end{center}
\end{table}

In Figure \ref{lcdm_mnu} we explore the resulting constraints on the neutrino properties. In all cases the dotted line is the WMAP 5-years only result, the dashed line is WMAP5+HST and the solid line is WMAP5+H. We find that the combination WMAP5+H constrains the sum of neutrino masses to be $\sum m_{\nu} < 0.48$ eV and $< 0.93$ eV at the 68\% and 95\% confidence levels, respectively, thus improving the WMAP-only constraints by 50\%. The constraints on the effective number of neutrino species is $N_{\rm rel} = 4.1^{+0.4}_{-0.9}\,(^{+1.1}_{-1.5})$ at the 68\% (95\%) confidence level and $N_{\rm rel} > 2.2 $ at better than the 99\% confidence level. For comparison, note that Ref.~\cite{de Bernardis:2007bu} obtained $N_{\rm rel} = 4.0 \pm 1.2$ at the 68\% confidence level and $N_{\rm rel} > 1.8$ at better than 99\% confidence level, from WMAP 3-years data combined with the same $H(z)$ measurements we use in this paper. Therefore, the improvement we obtain in the constraints of $N_{\rm rel}$, simply reflects the improvement of the CMB alone constraints in the WMAP 5-years data (compare e.g. the age constraints in table 1 of \cite{de Bernardis:2007bu} and our table 1). Note that our results are compatible (within the 1-2$\sigma$ regions) with the constraints on $N_{\rm eff}$ obtained~\cite{SimhaSteigman} from BBN alone or combined with other data sets, as well as have statistical errors of the same order (see table 1 from ~\cite{SimhaSteigman}).


\begin{figure}
\hspace{-0.55cm}\includegraphics[width=8cm, angle=0]{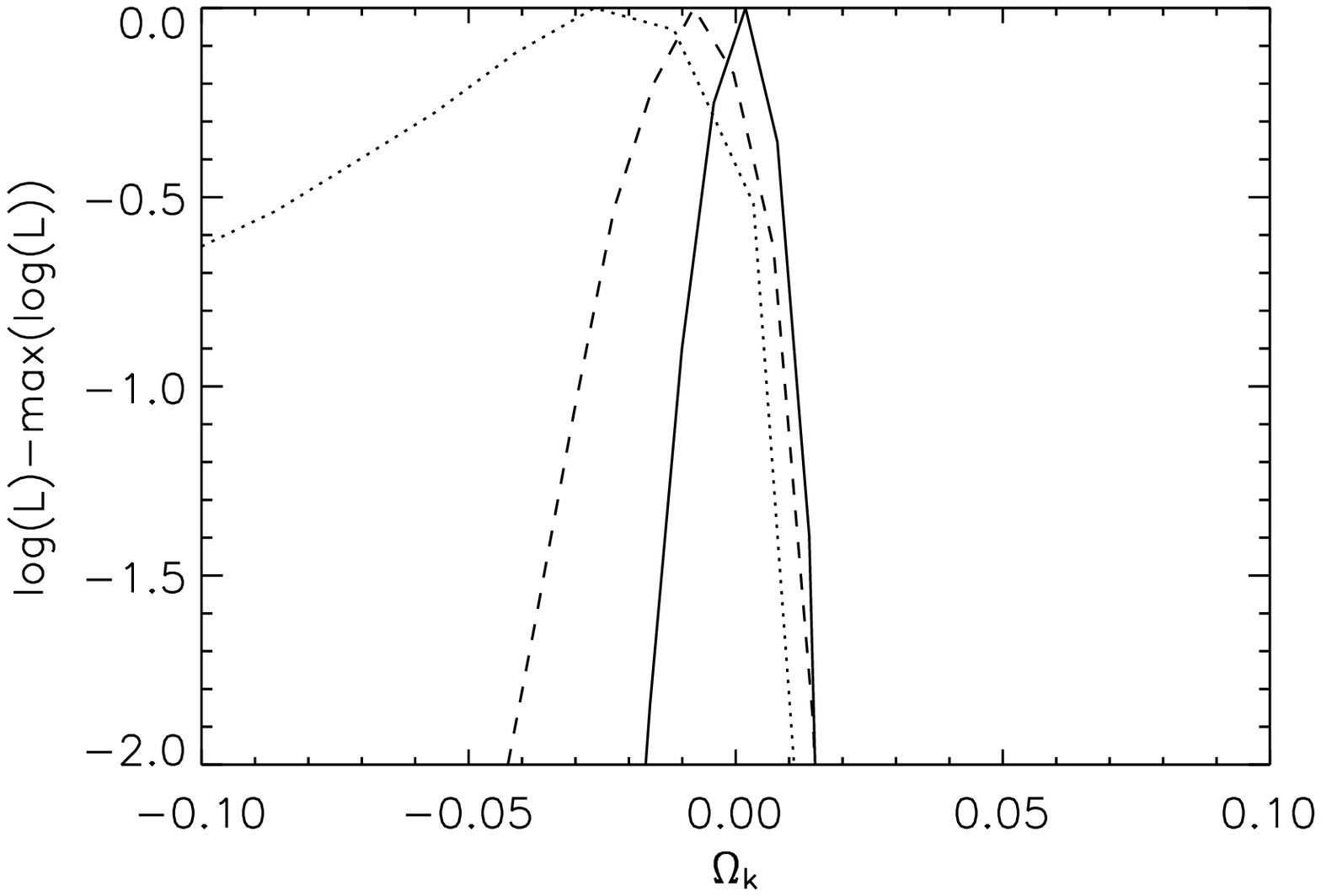}
\hspace{-0.1cm}\includegraphics[width=8cm, angle=0]{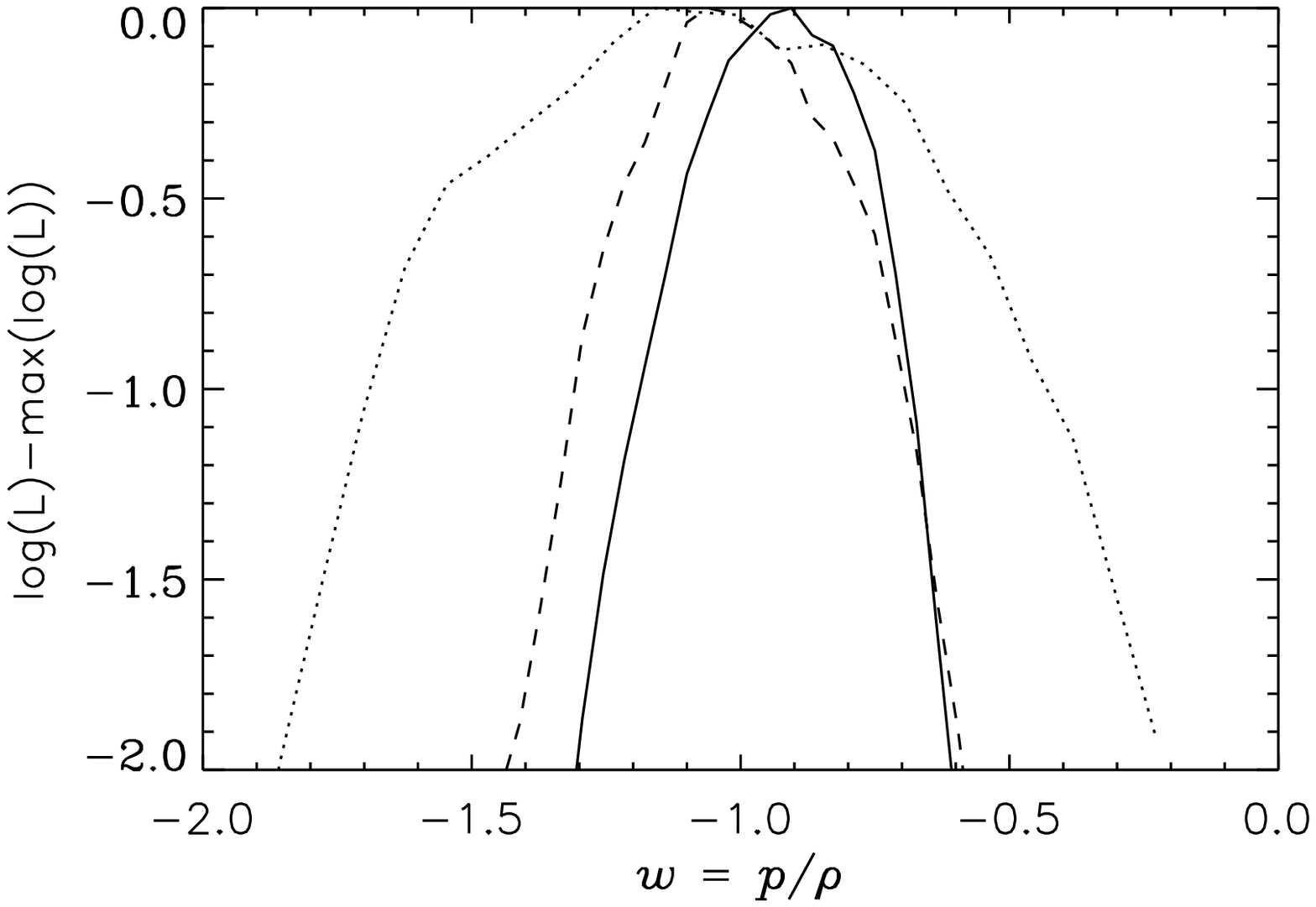}
\caption{\textit{Left}: Constraints on the curvature of a $\Lambda$CDM model from WMAP 5-years alone (dotted line), WMAP5+HST (dashed line) and WMAP5+H (solid line). With the $H(z)$ measurements, the curvature is constrained to 0.002 $\pm$ 0.006 ($\pm$ 0.014) at the $68\%$ ($95\%$) confidence level. The WMAP 5-years only line shows the well known geometric degeneracy. \textit{Right}: Constraints on the  dark energy equation of state parameter from WMAP 5-years alone (dotted line), WMAP5+HST (dashed line) and WMAP5+H (solid line). With the $H(z)$ measurements we obtain $w = -0.95 \pm 0.17\,(\pm\,0.32)$
at the 68\% (95\%) confidence level.}
\label{fig:olcdm}
\end{figure}


The left panel of Figure \ref{fig:olcdm} shows the constraints on the geometry of the Universe. The WMAP5+H combination yields  $\Omega_k=0.002 \pm 0.006\,(\pm\,0.014)$ at the 68\% (95\%) confidence levels, thus breaking the geometric degeneracy.

In  the right panel of Figure \ref{fig:olcdm}, we report the constraints on the dark energy equation of state parameter (asssumed constant). The WMAP5+H combination yields $w = -0.95 \pm 0.17\,(\pm\,0.32)$ at the  68\% (95\%) confidence level, which improves the WMAP 5-years only constraints by a factor $\sim 70\%$. While the WMAP 5-years constraint has a hard prior on the Hubble constant $H_0<100$ km/s/Mpc which imposes a lower limit on $w$, the WMAP5+HST and WMAP5+H combinations are insensitive to this prior.


In table \ref{tab:compare} we compare the WMAP5+H constraints  on deviations from the $\Lambda$CDM model, with those obtained by the combination WMAP 5-years with Baryon Acoustic Oscillation data (BAO) \cite{percivaletal07} and with Supernovae as obtained by \cite{komatsuwmap08}.

\begin{table}[htb]
\caption{\label{tab:compare} Cosmological constraints  at $68$\% ($95 \%$) c.l.\,\,on the extra parameters characterizing deviations of the standard $\Lambda$CDM model
, comparing their values as extracted from WMAP 5-years only, WMAP5+BAO+SN and WMAP5+H.}
\begin{center}
\begin{tabular}{|c||c|c|c|c|c|c|}
\hline
Parameter & WMAP 5-years only &  WMAP5+BAO+SN    & WMAP5+H\\
\hline
 & & &\\
$N_{\rm eff}$ & $> 2.3\,\, (95 \%)$ & $4.4^{+1.5}_{-1.5}\,^{(*)}$ & $4.10^{+0.37}_{-0.94} (^{+1.12}_{-1.50})$\\
& & &\\
$\sum m_\nu$ & $< 1.3$ eV (95 \%) & $< 0.61$ eV (95 \%) & $< 0.93$ eV (95 \%) \\
& & &\\
$w$ & $^{> -2.37}_{< -0.68}\,\, (95 \%)$ & $-0.972_{-0.060}^{+0.061} (_{-0.138}^{+0.112})$ & $-0.945_{-0.155}^{+0.194} (_{-0.350}^{+0.311})$\\
& & &\\
$\Omega_k$ & $^{< +0.017}_{> -0.063}\,\, (95 \%) $ & $-0.0052^{+0.0064}_{-0.0064} (^{+0.0137}_{-0.0123})$ & $0.002^{+0.0059}_{-0.0059} (^{+0.012}_{-0.018})$\\
& & & \\
\hline
\end{tabular}
\end{center}
$^{(*)}$with HST prior
\end{table}

\begin{figure}
\hspace{-0.0cm}\vspace{0.2cm}\includegraphics[width=7.5cm, height=6.2cm, angle=0]{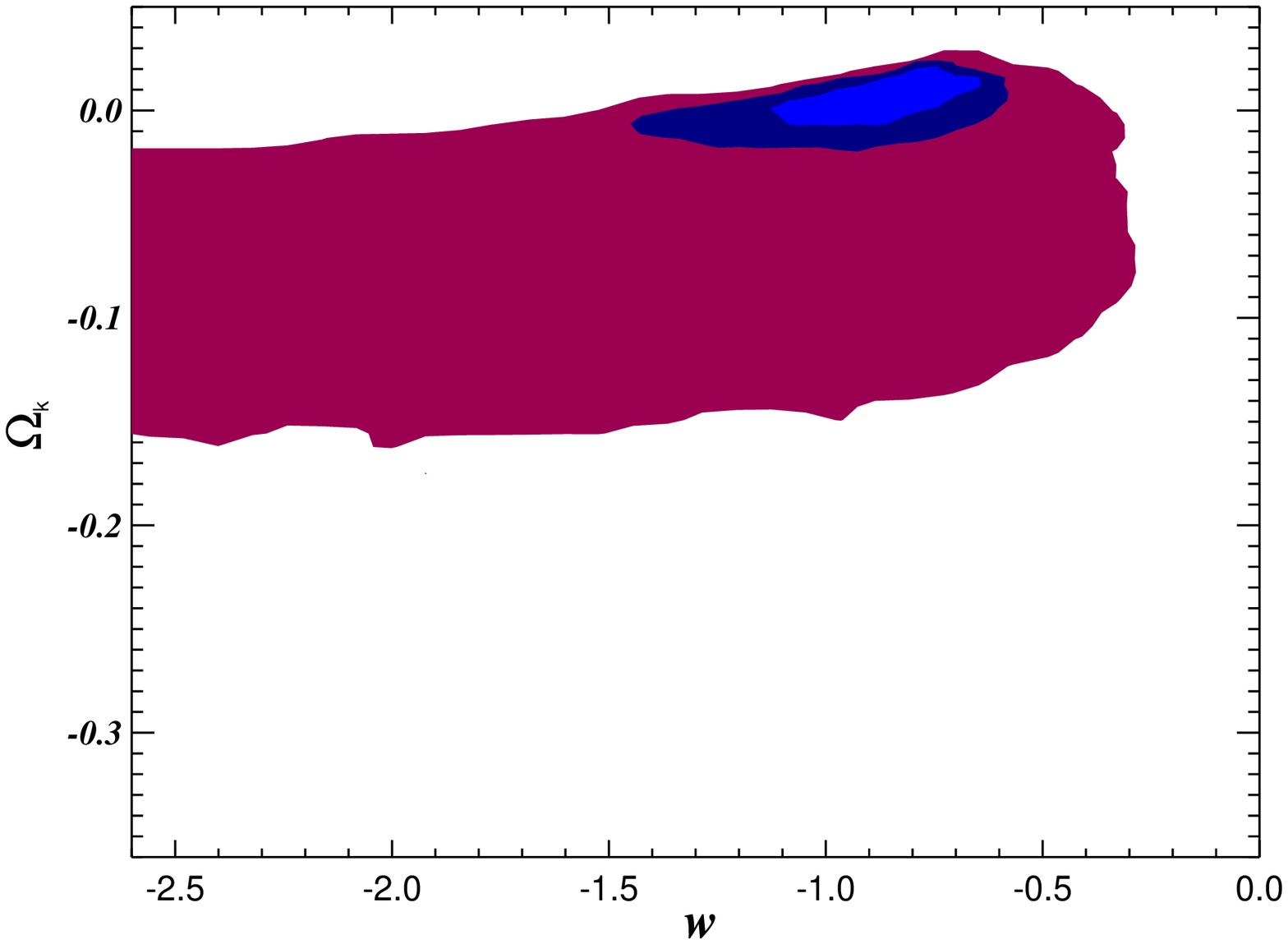}
\hspace{0.2cm}\includegraphics[width=7.5cm, height=5.8cm, angle=0]{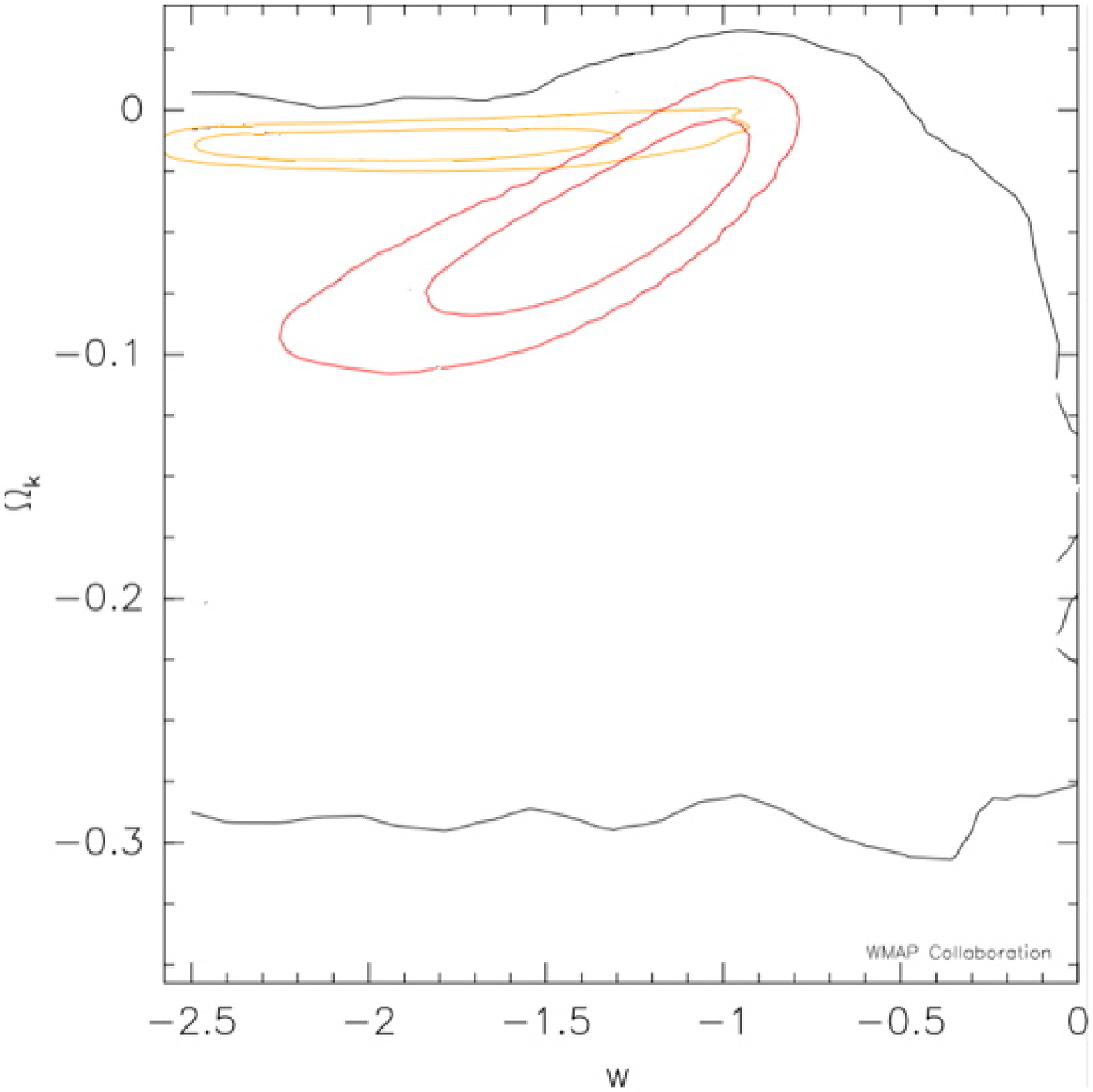}
\caption{\textit{Left}: Constraints in the $\Omega_k$-$w$ plane from WMAP 5-years alone (purple), WMAP5+H (blue 68\% and 95\% c.l.). For comparison we show (right) the wmap team's WMAP 5-years only constraints (black), WMAP5+BAO (yellow) and WMAP5+Supernovae (red) (see \cite{komatsuwmap08}). The differences in the WMAP 5-years only constraints are due to different choice of priors. Most notably, different boundaries on the $H_0$ prior are used: $0.4<h<1$ (left) vs $h<1$ (right), and on $w$ (on the left panel there is an additional  prior $w<-0.3$).}
\label{fig:owcdm}
\end{figure}

Finally, we  consider a model which deviates from the standard $\Lambda$CDM by two parameters: curvature is allowed to vary and the equation of state of dark energy (allowed to cluster), is assumed to be constant but not fixed to $w=-1$. When running the WMAP 5-years only Markov chain we use different priors from those used in \cite{dunkleywmap08}, the most important differences being on $h$ and $w$: we use $0.4<h<100$ and $-2.5<w<-0.3$, and a flat prior on the angular size distance to the last scattering surface rather than a flat prior on $\Omega_{\Lambda}$. In Figure \ref{fig:owcdm} we show  how the addition of $H(z)$ data helps break the  degeneracies. For comparison, on the right hand panel of Figure \ref{fig:owcdm} we show the constraints obtained by  \cite{komatsuwmap08} from the combination WMAP5+BAO and WMAP5+SN.
As already noted by e.g. \cite{weinberg70, polarski,ClarksonCortesBassett07}, measurements of $H(z)$ are crucial to break degeneracies between the curvature and dark energy properties.  
%

We conclude that the addition of Hubble parameter determinations at different redshifts, break the CMB-only degeneracies arising in models that allow deviations from the simple flat $\Lambda$CDM model. We find constraints on the number of effective neutrino species, the sum of neutrino masses, the curvature of the universe and the equation of state parameter for dark energy. These constraints are comparable to those obtained from the combination of WMAP 5-years with Supernovae and Baryon Acoustic Oscillations \cite{komatsuwmap08}. This ``concordance" approach shows that systematic errors in non-CMB data sets are smaller than the statistical errors and offers further support to the simple flat-$\Lambda$CDM model. Finally note that future BAO surveys will have the potential to constrain H(z) with \% accuracy in several z bins, i.e. they will constrain not only dark energy models but other deviations from the standard $\Lambda$CDM model, as explored in this paper.

\section*{Acknowledgments}

We would like to thank to Troels Haugboelle and Hiranya Peiris for useful comments on this work and help with MCMCs, and to A. Melchiorri for comments on the manuscript. We also thank the anonymous referee for comments that greatly improved the presentation of the material. We acknowledge the use of the Legacy Archive for Microwave Background Data Analysis (LAMBDA). Support for LAMBDA is provided by the NASA Office of Space Science. DGF is supported by a FPU contract, with ref.\,\,AP2005-1092. LV acknowledges support of FP7-PEOPLE-2007-4-3-IRGn202182 and CSIC I3 grant 200750I034. RJ is supported by CSIC I3 grant and FP7-PEOPLE-2007-IRG. This work was supported in part by the Spanish Ministry of Education and Science (MEC) through the Consolider Ingenio-2010 program, under project CSD2007-00060 Physics of the Accelerating Universe (PAU).

%
\section*{References}
 \bibliographystyle{JHEP}
 \providecommand{\href}[2]{#2}\begingroup\raggedright
\end{document}